\documentclass[aps,floatfix,superscriptaddress,twocolumn,tightenlines,amsmath,amssymb,nofootinbib,10pt]{revtex4-1}

\usepackage[T1]{fontenc}
\usepackage{graphicx}
\usepackage{amssymb}
\usepackage{xcolor}
\usepackage{amsthm}
\usepackage{psfrag}
\usepackage{ifsym}
\usepackage{dsfont}
\usepackage{enumerate}
\usepackage{amsfonts}
\usepackage{multirow} 
\usepackage{amsmath}
\usepackage{hyperref}
\hypersetup{colorlinks=true,linkcolor=red,citecolor=blue}
\usepackage[separate-uncertainty=true]{siunitx}
\usepackage[normalem]{ulem}

\newcommand{\bra}[1] {\langle #1 |}
\newcommand{\ket}[1] {| #1 \rangle}

\newcommand{\ketbra}[1]{ | #1 \rangle\!\langle #1 |}

\newcommand{\expec}[1]{\left\langle #1 \right\rangle}
\newcommand{\one}{\leavevmode\hbox{\small1\normalsize\kern-.33em1}}

\newcommand{\id}{\mathds{1}}

\begin{document}
\title{Experimental test of local observer-independence}

\author{Massimiliano Proietti}
\affiliation{
Scottish Universities Physics Alliance (SUPA), Institute of Photonics and Quantum Sciences, School of Engineering and Physical Sciences, Heriot-Watt University, Edinburgh EH14 4AS, UK.}

\author{Alexander Pickston}
\affiliation{
Scottish Universities Physics Alliance (SUPA), Institute of Photonics and Quantum Sciences, School of Engineering and Physical Sciences, Heriot-Watt University, Edinburgh EH14 4AS, UK.}

\author{Francesco Graffitti}
\affiliation{
Scottish Universities Physics Alliance (SUPA), Institute of Photonics and Quantum Sciences, School of Engineering and Physical Sciences, Heriot-Watt University, Edinburgh EH14 4AS, UK.}

\author{Peter Barrow}
\affiliation{
Scottish Universities Physics Alliance (SUPA), Institute of Photonics and Quantum Sciences, School of Engineering and Physical Sciences, Heriot-Watt University, Edinburgh EH14 4AS, UK.}

\author{Dmytro Kundys}
\affiliation{
Scottish Universities Physics Alliance (SUPA), Institute of Photonics and Quantum Sciences, School of Engineering and Physical Sciences, Heriot-Watt University, Edinburgh EH14 4AS, UK.}

\author{Cyril Branciard}
\affiliation{Universit{\'e} Grenoble Alpes, CNRS, Grenoble INP, Institut N{\'e}el, 38000 Grenoble, France.}

\author{Martin Ringbauer}
\affiliation{
Scottish Universities Physics Alliance (SUPA), Institute of Photonics and Quantum Sciences, School of Engineering and Physical Sciences, Heriot-Watt University, Edinburgh EH14 4AS, UK.}
\affiliation{Institut f{\"u}r Experimentalphysik, Universit{\"a}t Innsbruck, 6020 Innsbruck, Austria.}

\author{Alessandro Fedrizzi}
\affiliation{
Scottish Universities Physics Alliance (SUPA), Institute of Photonics and Quantum Sciences, School of Engineering and Physical Sciences, Heriot-Watt University, Edinburgh EH14 4AS, UK.}

\begin{abstract}
The scientific method relies on facts, established through repeated measurements and agreed upon universally, independently of who observed them. In quantum mechanics, the objectivity of observations is not so clear, most dramatically exposed in Eugene Wigner's eponymous thought experiment where two observers can experience seemingly different realities. The question whether these realities can be reconciled in an observer-independent way has long remained inaccessible to empirical investigation, until recent no-go-theorems constructed an extended Wigner's friend scenario with four observers that allows us to put it to the test. In a state-of-the-art 6-photon experiment, we realise this extended Wigner's friend scenario, experimentally violating the associated Bell-type inequality by 5 standard deviations. If one holds fast to the assumptions of locality and free-choice, this result implies that quantum theory should be interpreted in an observer-dependent way.
\end{abstract}

\maketitle

\paragraph*{Introduction.---}
The observer's role as final arbiter of universal facts~\cite{Popper1992} was imperilled by the advent of 20$^{\textrm{th}}$ century science. In relativity, previously absolute observations are now relative to moving reference frames; in quantum theory, all physical processes are continuous and deterministic, except for observations, which are proclaimed to be instantaneous and probabilistic. This fundamental conflict in quantum theory is known as the measurement problem, and it originates because the theory does not provide a clear cut between a process being a measurement or just another unitary physical interaction.

\begin{figure}[t!]
  \begin{center}
\includegraphics[width=0.95\columnwidth]{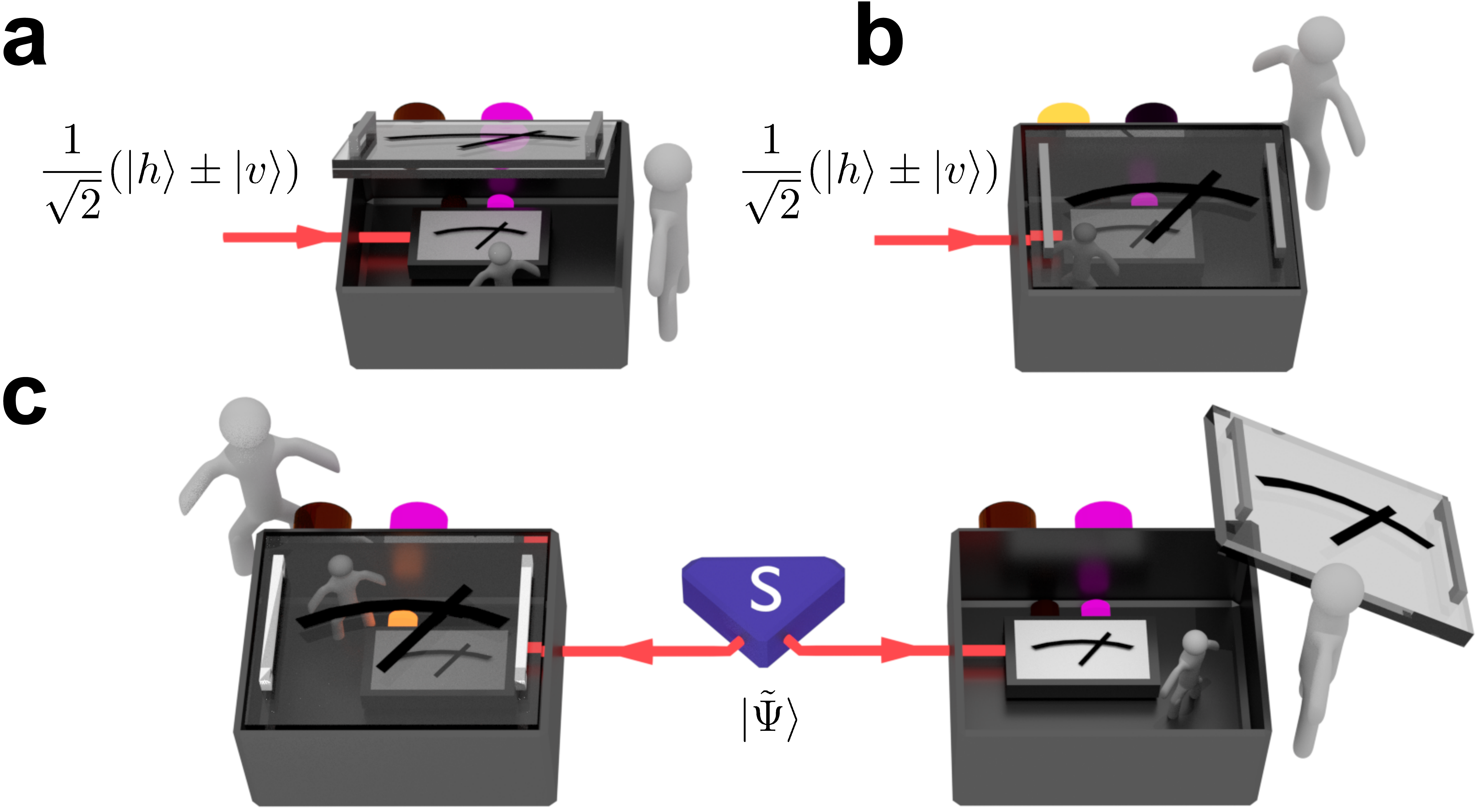}
  \end{center}
\caption{\textbf{Wigner's friend experiment.}
\textbf{a} A quantum system in an equal superposition of two possible states is measured by Wigner's friend (inside the box). According to quantum theory, in each run she will randomly obtain one of the two possible measurement outcomes. This can indeed be verified by directly looking into her lab and reading which result she recorded. \textbf{b} From outside the closed laboratory, however, Wigner must describe his friend and her quantum system as a joint entangled state. Wigner can also verify this state assignment through an interference experiment, concluding that his friend cannot have seen a definite outcome in the first place.
\textbf{c} We consider an extended version of that experiment, where an entangled state is sent to two different laboratories, each involving an experimenter and their friend.}
\label{fig:Motivation}
\end{figure}

\begin{figure*}[t!]
  \begin{center}
\includegraphics[width=0.95\textwidth]{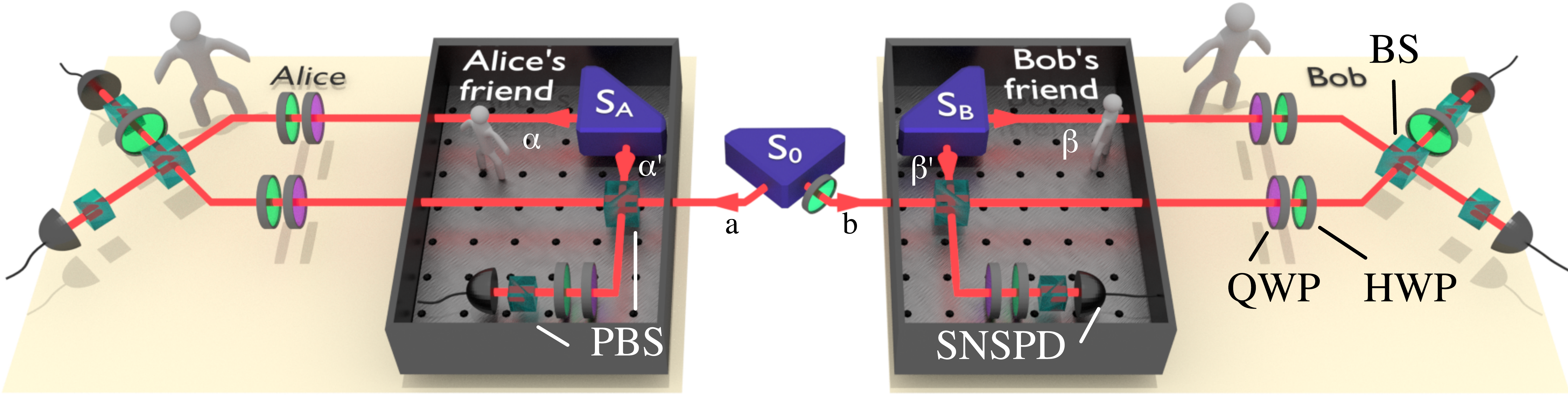}
  \end{center}
\caption{\textbf{Experimental setup.}
Pairs of entangled photons from the source $S_0$, in modes $a$ and $b$, respectively, are distributed to Alice's and Bob's friends, who locally measure their respective photon in the $\{h,v\}$-basis using entangled sources $S_A, S_B$ and type-I fusion gates. These use nonclassical interference on a polarising beam splitter (PBS) together with a set of half-wave (HWP) and quarter-wave plates (QWP). The photons in modes $\alpha'$ and $\beta'$ are detected using superconducting nanowire single-photon detectors (SNSPD) to herald the successful measurement, while the photons in modes $\alpha$ and $\beta$ record the friends' measurement results. Alice (Bob) then either performs a Bell-state measurement via non-classical interference on a $50/50$ beam splitter (BS) on modes $a$ and $\alpha$ ($b$ and $\beta$) to measure $A_1$ ($B_1$) and establish her (his) own fact, or removes the BS to measure $A_0$ ($B_0$), to infer the fact recorded by their respective friend; see Supplementary Materials for details.}
\label{fig:Setup}
\end{figure*}

This is best illustrated in the seminal ``Wigner's friend'' thought experiment~\cite{Wigner1961}, whose far-reaching implications are only starting to become clear~\cite{Brukner2015MeasurementProblem,Brukner2018,Frauchiger2016}. Consider a single photon in a superposition of horizontal $\ket{h}$ and vertical polarisation $\ket{v}$, measured in the $\{\ket{h},\ket{v}\}$-basis by an observer---Wigner's friend---in an isolated lab, see Figs.~\hyperref[fig:Motivation]{\ref*{fig:Motivation}\textbf{a}} and \hyperref[fig:Motivation]{\textbf{b}}. According to quantum theory, the friend randomly observes one of the two possible outcomes in every run of the experiment. The friend's record, $h$ or $v$, can be stored in one of two possible orthogonal states of some physical memory, labeled either $\ket{\text{``photon is \textit{h}''}}$ or $\ket{\text{``photon is \textit{v}''}}$, and constitutes a ``fact'' from the friend's point of view. Wigner, who observes the isolated laboratory from the outside, has no information about his friend's measurement outcome. According to quantum theory Wigner must describe the friend's measurement as a unitary interaction that leaves the photon and friend's record in the entangled state (with implicit tensor products):
\begin{align}
&\frac{1}{\sqrt{2}}(\ket{h}\pm\ket{v}) \nonumber \\
&\to \frac{1}{\sqrt{2}} \big(\ket{h}\ket{\text{``photon is \textit{h}''}} \pm \ket{v}\ket{\text{``photon is \textit{v}''}} \big) \nonumber\\
& \qquad \qquad \qquad \qquad =: \ket{\Phi^\pm_{\text{photon/record}}}.
\label{eq:wigner}    
\end{align}
Wigner can now perform an interference experiment in an entangled basis containing the states of Eq.~\eqref{eq:wigner} to verify that the photon and his friend's record are indeed in a superposition---a ``fact'' from his point of view. From this fact, Wigner concludes that his friend cannot have recorded a definite outcome. Concurrently however, the friend does always record a definite outcome, which suggests that the original superposition was destroyed and Wigner should not observe any interference. The friend can even tell Wigner that she recorded a definite outcome (without revealing the result), yet Wigner and his friend's respective descriptions remain unchanged~\cite{Deutsch1985}.
This calls into question the objective status of the facts established by the two observers. Can one reconcile their different records, or are they fundamentally incompatible---so that they cannot be considered objective, observer-independent ``facts of the world''~\cite{Brukner2015MeasurementProblem,Brukner2018}?

% =========================================================
% Theory background
% =========================================================
It was recently shown~\cite{Brukner2018} that this question can be addressed formally, by considering an extension of the Wigner's friend scenario as follows. Consider a pair of physical systems, shared between two separate laboratories controlled by Alice and Bob, respectively, see Fig.~\hyperref[fig:Motivation]{\ref*{fig:Motivation}\textbf{c}}. Inside these laboratories, Alice's friend and Bob's friend measure their respective system non-destructively and record the outcomes in some memory. Outside these laboratories, in each run of the experiment Alice and Bob can choose to either measure the state of their friend's record---i.e.\ to attest the ``facts'' established by their friend, and whose results define the random variables $A_0$ (for Alice's friend) and $B_0$ (for Bob's friend); or to jointly measure the friend's record and the system held by the friend---to establish their own ``facts'', defining variables $A_1$ (for Alice) and $B_1$ (for Bob). After comparing their results, Alice and Bob can estimate the probability distributions $P(A_x,B_y)$ for all four combinations of $x,y = 0,1$. As in the original Wigner's friend Gedankenexperiment, the facts $A_1, B_1$ attributed to Alice and Bob and $A_0, B_0$ attributed to their friends' measurements may be inconsistent.

This raises the question whether a more general framework exists in which all observers can reconcile their recorded facts. We shall call this assumption O, \emph{observer-independent facts}, stating that a record or piece of information obtained from a measurement should be a ``fact of the world'' that all observers can agree on---and that such ``facts'' take definite values even if not all are ``co-measured''~\cite{Healey2019,Brukner2019}. Under the additional assumptions of \emph{locality}~(L), that Alice and Bob's choices do not influence each others' outcome, and free-choice~(F), that Alice and Bob can freely choose their measurements $A_0, A_1$ and $B_0, B_1$, it should then be possible to construct a single probability distribution $P(A_0,A_1,B_0,B_1)$ for the four individual facts under consideration, whose marginals match the probabilities $P(A_x,B_y)$~\cite{Brukner2015MeasurementProblem,Brukner2018}.

Any joint probability distribution satisfying these assumptions must then satisfy Bell inequalities~\cite{Fine1982}. More specifically, when the variables $A_x, B_y$ take values $a,b\in\{-1,+1\}$, then the average values $\langle A_xB_y \rangle = \sum_{a,b}a b P(A_x = a,B_y=b)$ must obey the Clauser-Horne-Shimony-Holt inequality~\cite{CHSH1969}:
\begin{equation}
 S  = \langle A_1B_1 \rangle + \langle A_1B_0 \rangle + \langle A_0B_1 \rangle - \langle A_0B_0 \rangle \leq 2 .
\label{eq:CHSH} 
\end{equation}
As shown in Refs.~\cite{Brukner2015MeasurementProblem,Brukner2018}, a violation of the inequality above is, however, possible in a physical world described by quantum theory. Such a violation would demonstrate that the observed probability distributions $P(A_x,B_y)$ are incompatible with assumptions F, L, and O. Therefore, if we accept F and L, it follows that the pieces of information corresponding to facts established by Alice, Bob, and their friends cannot coexist within a single, observer-independent framework~\cite{Brukner2015MeasurementProblem,Brukner2018}. Notably this is the case even though Alice and Bob can acknowledge the occurrence of a definite outcome in their friend's closed laboratory.

We note that, although Bell's mathematical machinery~\cite{bell_aspect_2004} is used to show the result, the set of assumptions considered here---and therefore the conclusions that can be drawn from a violation of inequality~\eqref{eq:CHSH}---are different from those in standard Bell tests. In fact, while they share assumptions L and F, the third assumption of predetermination (PD) in the original Bell theorem~\cite{Bell1964}, for instance, differs from our assumption O in that it is only concerned with the deterministic (or otherwise) nature of measurement outcomes, not with their objectivity as in O. A Bell test is indifferent to the observables used and the underlying system, such that any violation suffices to rule out the conjunction of L, F and PD. In contrast, a Bell-Wigner test is based on very specific observables that satisfy the definition of an observation given below and thus represent facts relative to different observers. Formally, any Bell-Wigner violation implies a Bell-violation, but not the other way round.

Before we describe our experiment in which we test and indeed violate inequality~\eqref{eq:CHSH}, let us first clarify our notion of an observer. Formally, an observation is the act of extracting and storing information about an observed system. Accordingly, we define an observer as any physical system that can extract information from another system by means of some interaction, and store that information in a physical memory.

Such an observer can establish ``facts'', to which we assign the value recorded in their memory. Notably, the formalism of quantum mechanics does not make a distinction between large (even conscious) and small physical systems, which is sometimes referred to as universality. Hence, our definition covers human observers, as well as more commonly used non-conscious observers such as (classical or quantum) computers and other measurement devices---even the simplest possible ones, as long as they satisfy the above requirements. We note that the no-go theorem formulated in~\cite{Frauchiger2016} requires observers to be ``agents'', who ``use'' quantum theory to make predictions based on the measurement outcomes. In contrast, for the no-go theorem we tested here~\cite{Brukner2018} it is sufficient that they perform a measurement and record the outcome. The enhanced capabilities required of agents were recently discussed in \cite{Baumann2019}.

% =========================================================
% Experiment
% =========================================================
\paragraph*{Results.---}
Our experiment makes use of three carefully designed~\cite{GraffittiOptica, GraffittiPRA} sources $S_0, S_A$ and $S_B$, see Fig.~\ref{fig:Setup}, which generate pairs of 1550~nm single photons, entangled in the polarisation degree of freedom~\cite{Jin14} in the state $\ket{\Psi^{-}}=\left(\ket{h}\ket{v}-\ket{v}\ket{h}\right)/\sqrt{2}$. We confirmed the almost ideal quality of the prepared states via quantum state tomography, with typical fidelity $\mathcal{F}=99.62^{+0.01}_{-0.04}\%$, purity $\mathcal{P}=99.34^{+0.01}_{-0.09}\%$
and entanglement as measured by the concurrence $\mathcal{C}=99.38^{+0.02}_{-0.10}\%$, see Supplementary Materials for details. The photon pair from source $S_0$ is rotated to
\begin{equation}
\ket{\widetilde\Psi}= \id \otimes U_{\frac{7\pi}{16}} \, \ket{\Psi^{-}},
\label{Eq:ResourceState}
\end{equation}
using a half-wave plate at an angle $7\pi/16$, given by $U_{\frac{7\pi}{16}} = \cos (\frac{7\pi}{8}) \, \sigma_z + \sin (\frac{7\pi}{8}) \, \sigma_x$ (where $\id$ is the identity, $\sigma_z, \sigma_x$ are the Pauli operators).
This state maximises the violation of inequality~\eqref{eq:CHSH} for our choice of measurement settings, see Eq.~\eqref{Eq:Observables}.

Source $S_0$ provides the quantum systems on which Alice's and Bob's friends perform their measurements. Recalling the above definition of an observer, we employ the entangled photon pairs from sources $S_A$ and $S_B$ as the physical systems which, through interaction in a type-I fusion gate~\cite{Browne2005,PanFusionGate} between modes $a$, $\alpha'$ and $b$, $\beta'$ respectively (see Fig.~\ref{fig:Setup}), are able to extract information and thereby establish their own facts. When successful, the fusion gate realises a non-destructive polarisation measurement of a photon from $S_0$ in the $\{\ket{h},\ket{v}\}$-basis, whose results $\ket{\text{``photon is \textit{h}''}}$ or $\ket{\text{``photon is \textit{v}''}}$ represent the friend's record. Via the ancillary entanglement, the extracted information is then stored in the polarisation state of the other photon from $S_A$ ($S_B$)---in mode $\alpha$ $(\beta)$---which acts as a memory, while the photon in mode $\alpha'$ $(\beta')$ is absorbed in a single photon counter to herald the success of the measurement (see Supplementary Materials for details). Note that this detection could be delayed until the end of the experiment as it carries no information about the measurement outcome, akin to the observer in the box communicating the fact that an observation took place~\cite{Brukner2015MeasurementProblem,Brukner2018}. From Alice's and Bob's perspective, the yet undetected photons from $S_0, S_A$, and $S_B$ are now in a joint 4-photon entangled state, see Eq.~\eqref{eq:4photon_state} in the Supplementary Materials.

\begin{figure}[h!]
  \begin{center}
\includegraphics[width=0.95\columnwidth]{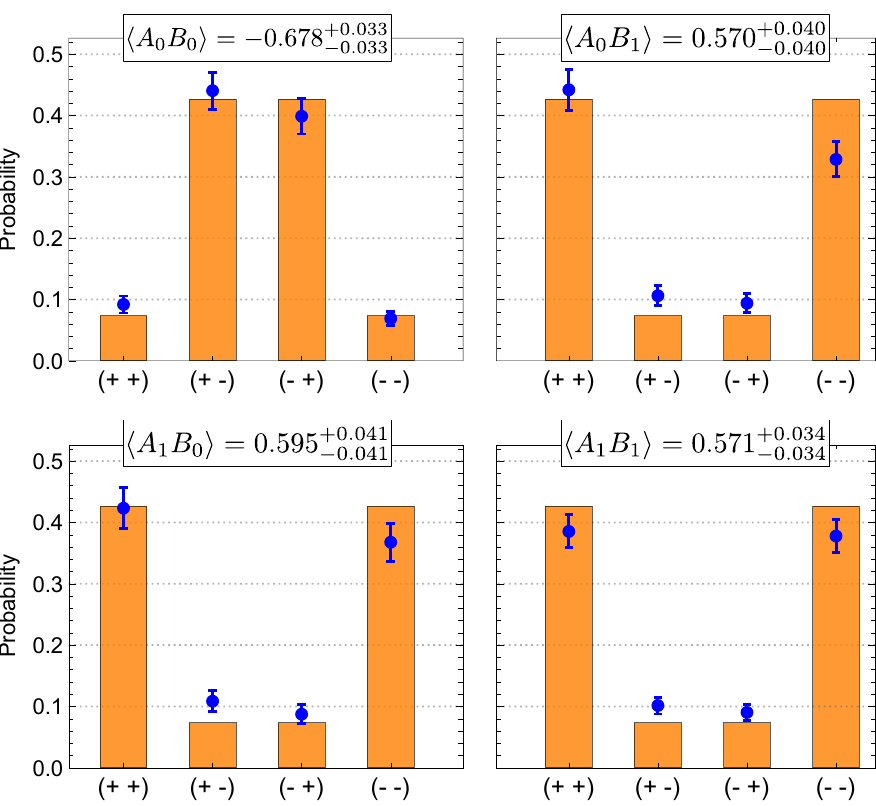}
  \end{center}
\caption{\textbf{Experimental data}. The outcome probabilities comprising each of the four expectation values $\expec{A_0B_0}$, $\expec{A_0B_1}$, $\expec{A_1B_0}$, $\expec{A_1B_1}$ are obtained from the measured 6-fold coincidence events for each set of $4 \times 4$ eigenvectors during a fixed time window. Shown here, are only the data corresponding to non-zero eigenvalues labelled on the horizontal axes $+$ and $-$ for $+1$ and $-1$, respectively, with the full data shown in the Supplementary Materials. The theoretical predictions are shown as orange bars, and each measured expectation value is given above the corresponding sub-figure. Uncertainties on the latter and error bars on the data represent $1\sigma$ statistical confidence intervals assuming Poissonian counting statistics (see Supplementary Materials).}  
\label{fig:Results}
\end{figure}

To test inequality~\eqref{eq:CHSH}, Alice and Bob then measure the following observables on their respective joint photon / friend's record systems:
\begin{align}
A_0 \!=\! B_0 & \!=\! \id \otimes \big(\ketbra{\text{``photon is \textit{h}''}} \notag \\
& \hspace{12mm} - \ketbra{\text{``photon is \textit{v}''}}\big), \notag \\[2mm]
A_1 \!=\! B_1 & \!=\! \ketbra{\Phi^+_{\text{photon/record}}} \notag \\
& \quad - \ketbra{\Phi^-_{\text{photon/record}}}.
\label{Eq:Observables}
\end{align}
(with $\ket{\Phi^\pm_{\text{photon/record}}}$ as defined in Eq.~\eqref{eq:wigner}). The observables $A_0$ and $B_0$ directly unveil the records established by Alice's and Bob's friend, respectively. The observables $A_1$ and $B_1$, on the other hand, correspond to Alice's and Bob's joint measurements on their friend's photon and record, and define their own facts in the same way as Wigner in the original thought experiment confirms his entangled state assignment.

% =========================================================
% Results
% =========================================================
We estimate the four average values $\langle A_x B_y \rangle$ in inequality~\eqref{eq:CHSH} via projection onto each of the $4 \times 4$ eigenstates of the observables $A_x$ and $B_y$, see Supplementary Materials for details. For the corresponding 64 settings we collect $1794$ six-photon coincidence events over a total measurement time of 360 hours, from which we calculate the probabilities shown in Fig.~\ref{fig:Results}.
We achieve a value of $S_{\textrm{exp}} = 2.416^{+0.075}_{-0.075}$, thus violating inequality~\eqref{eq:CHSH} by more than $5$ standard deviations. This value is primarily limited by the higher-order photon emissions from our probabilistic photon sources. Statistical uncertainties are independently estimated using an error propagation approach and a Monte-Carlo method. Details are discussed in the Supplementary Materials. 

% =========================================================
% Discussion
% =========================================================
\paragraph*{Discussion.---} 
In principle, ``Bell-Wigner tests'' like ours are subject to similar loopholes as tests of conventional Bell inequalities~\cite{LarssonReview}. To address the detection and space-time loopholes, we make the physically reasonable assumption of fair sampling and rely on the empirical absence of signalling between our measurement devices (which experimentally we verified to be in agreement with the expectation from Poissonian statistics), respectively. Another loophole may arise if the observables $A_0, B_0$ that are measured in practice do not strictly correspond to a measurement of the friends' memories. Here we assume (with reasonable confidence, up to negligible experimental deviations) that the measured observables indeed factorise as in Eq.~\eqref{Eq:Observables}, with the identity on the photon system, so that the above interpretation for $A_0$, $B_0$ can be trusted. As discussed in the Supplementary materials, closing all loopholes in full will be considerably more challenging than for Bell tests. 

One might further be tempted to deny our photonic memories the status of ``observer''. This, however, would require a convincing revision of our minimal definition of what qualifies as an observer, which typically comes at the cost of introducing new physics that is not described by standard quantum theory. Eugene Wigner, for example, argued that the disagreement with his hypothetical friend could not arise due to a supposed impossibility for conscious observers to be in a superposition state~\cite{Wigner1961}. However, the lack of objectivity revealed by a Bell-Wigner test does not arise in anyone's consciousness, but between the recorded facts. Since quantum theory does not distinguish between information recorded in a microscopic system (such as our photonic memory) and in a macroscopic system the conclusions are the same for both: the measurement records are in conflict regardless of the size or complexity of the observer that records them. Implementing the experiment with more complex observers would not necessarily lead to new insights into the specific issue of observer-independence in quantum theory. It would however serve to show that quantum mechanics still holds at larger scales, ruling out alternative (collapse) models~\cite{GRW1986}. However, this is not the point of a Bell-Wigner test---less demanding experiments could show that. 

Modulo the potential loopholes and accepting the photons' status as observers, the violation of inequality~\eqref{eq:CHSH} implies that at least one of the three assumptions of free choice, locality, and observer-independent facts must fail. The related no-go theorem by Frauchiger \& Renner~\cite{Frauchiger2016} rests on different assumptions which do not explicitly include locality. While the precise interpretation of Ref.~\cite{Frauchiger2016} within non-local theories is under debate~\cite{Lazarovici2019}, it seems that
abandoning free choice and locality might not resolve the contradiction~\cite{Frauchiger2016}. A compelling way to accommodate our result is then to proclaim that ``facts of the world'' can only be established by a privileged observer---e.g., one that would have access to the ``global wavefunction'' in the many worlds interpretation~\cite{Everett1957} or Bohmian mechanics~\cite{Bohm1952}. Another option is to give up observer independence completely by considering facts only relative to observers~\cite{Rovelli1996}, or by adopting an interpretation such as QBism, where quantum mechanics is just a tool that captures an agent's subjective prediction of future measurement outcomes~\cite{Fuchs2017}. This choice, however, requires us to embrace the possibility that different observers irreconcilably disagree about what happened in an experiment. A further interesting question is whether the conclusions drawn from Bell-, or Bell-Wigner tests change under relativistic conditions with non-inertial observers~\cite{Durham2019}.

\bibliography{bib}
\bibliographystyle{apsrev}

\bigskip

\textbf{Acknowledgements} 
We thank {\v C}. Brukner, R. Renner and F. Shahandeh for useful discussions. This work was supported by the UK Engineering and Physical Sciences Research Council (grant number EP/N002962/1, EP/L015110/1) and the French National Research Agency (grant number ANR-13-PDOC-0026). This project has received funding from the European Union's Horizon 2020 research and innovation programme under the Marie Sk\l{}odowska-Curie grant agreement No 801110 and the Austrian Federal Ministry of Education, Science and Research (BMBWF).\\

\textbf{Correspondence} Correspondence and requests for materials should be addressed to AF (email: a.fedrizzi@hw.ac.uk).\\

\newpage
\appendix
\clearpage
\renewcommand{\theequation}{S\arabic{equation}}
\renewcommand{\thefigure}{S\arabic{figure}}
\renewcommand{\thetable}{\Roman{table}}
\renewcommand{\thesection}{S\Roman{section}}
\setcounter{equation}{0}
\setcounter{figure}{0}

\section*{Supplementary materials}

\begin{figure*}[t!]
  \begin{center}
\includegraphics[width=0.95\textwidth]{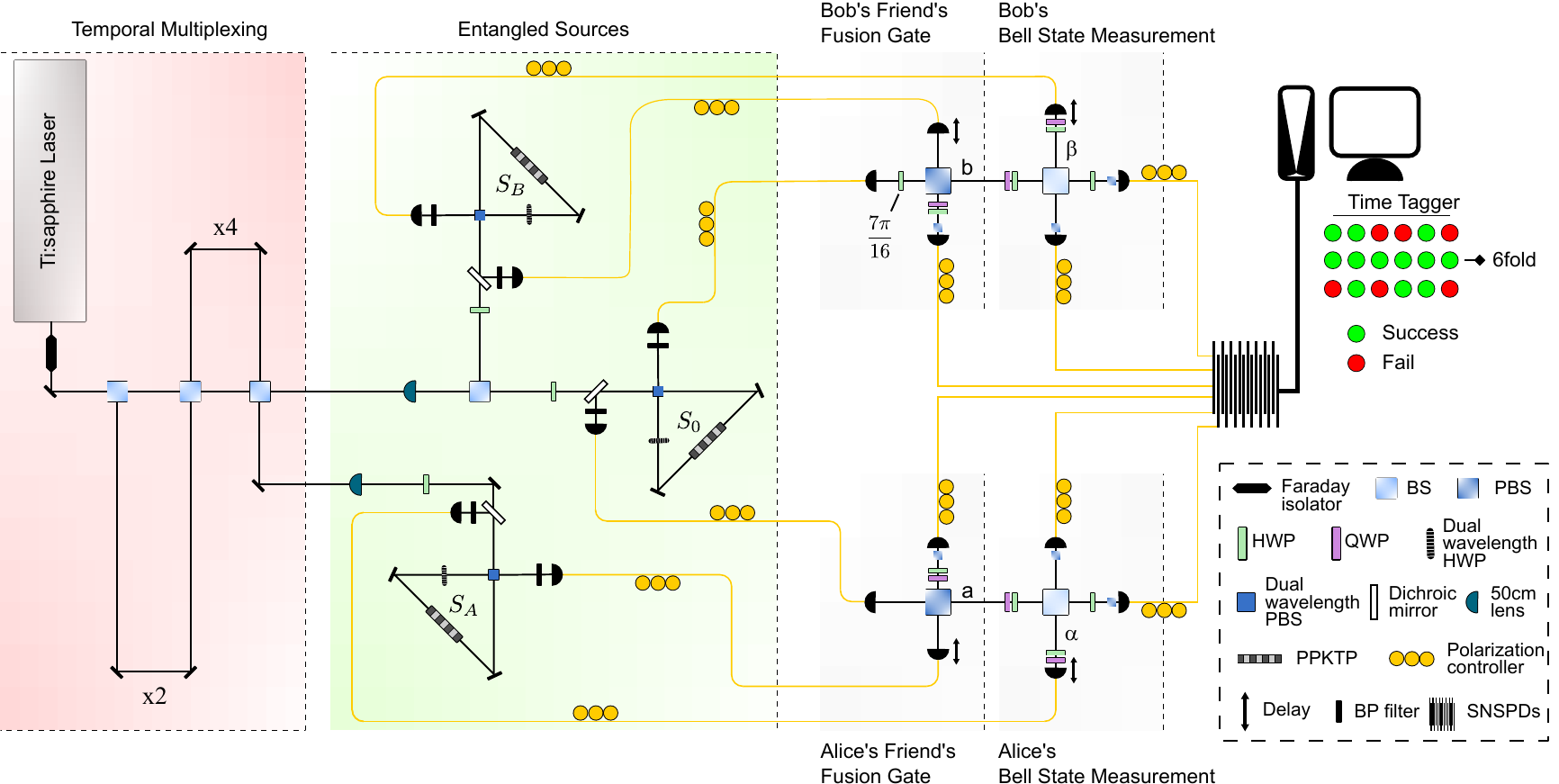}
  \end{center}
\caption{\textbf{Detailed experimental setup.} The Ti:sapphire laser beam is protected from back-reflections by a Faraday isolator and spatially filtered using a short single-mode fibre (not shown). The laser beam is then temporally multiplexed to effectively quadruple the pulse rate. The pump is then delivered to three Sagnac-interferometer sources to create polarisation entangled photon pairs. The outputs of each source are coupled to single-mode fibres and delivered to the measurement stages. Fibre polarisation controllers are used to maintain the polarisation states of the photons during transport. The three entangled pairs are then subject to two fusion gates, where temporal mode matching is achieved by employing physical delays as indicated. One photon at each measurement stage acts as a heralding signal for the success of the fusion gate, while the other two are subject to a Bell-state measurement on a 50/50 beam splitter, or to a direct measurement without the BS (for $A_0,B_0$), followed by projection onto orthogonal polarisations. Finally, all six photons are fibre-coupled and detected by the SNSPDs whose detection is processed by a classical computer to find 6-photon coincidence events.}
  \label{fig.MethodsSetup}
 \end{figure*}

\paragraph*{Setup details.--- }
A \SI{775}{nm}, \SI{1.6}{ps}-pulsed Ti:Sapphire laser is focused into a \SI{22}{mm} periodically-poled KTP crystal in a Sagnac-type interferometer~\cite{Fedrizzi2007Sagnac}, where it generates pairs of \SI{1550}{nm} single photons through collinear type-II parametric down-conversion. The \SI{80}{MHz} repetition rate of the pump laser is quadrupled through temporal multiplexing~\cite{Broome2011} in order to suppress higher-order emissions, see Fig.~\ref{fig.MethodsSetup}. We thereby achieve a signal-to-noise ratio (i.e.\ photon pairs vs.\ higher-order contributions) of $140\pm 10$ in each photon source, generating $\sim 8000$ photon pairs mW$^{-1}$ s$^{-1}$ with a typical heralding efficiency $\eta=(cc/\sqrt{s_{1}s_{2}})$ of $\sim 50\%$, where $cc$ are the number of coincidence counts, and $s_1, s_2$ are the numbers of singles in the first and second output respectively. Single photons pass through \SI{3}{nm} band-pass (BP) filters to guarantee high spectral purity, and are detected with superconducting nano-wire single-photon detectors (SNSPDs) with a detection efficiency of $\sim 80\%$. Detector clicks are time-tagged using a field-programmable gate-array and processed to detect coincidences within a temporal window of \SI{1}{ns}.

To benchmark the three required 2-qubit states we perform maximum-likelihood quantum state tomography directly at each source. From the reconstructed density matrices we compute the fidelity, concurrence and purity quoted in the main text. Further transmission of the photon pairs to the fusion gates slightly degrade the fidelities of the three entangled pairs, to $\mathcal{F}_{0}=98.79^{+0.03}_{-0.03}\%$, $\mathcal{F}_{A}=98.70^{+0.03}_{-0.03}\%$, $\mathcal{F}_{B}=98.59^{+0.03}_{-0.03}\%$ for sources $S_0, S_A$ and $S_B$ respectively, see Fig.~\ref{fig:Setup}. This indicates that the optical circuit preserves the excellent quality of the initial states.

 \begin{figure*}[t!]
  \begin{center}
\includegraphics[width=0.95\textwidth]{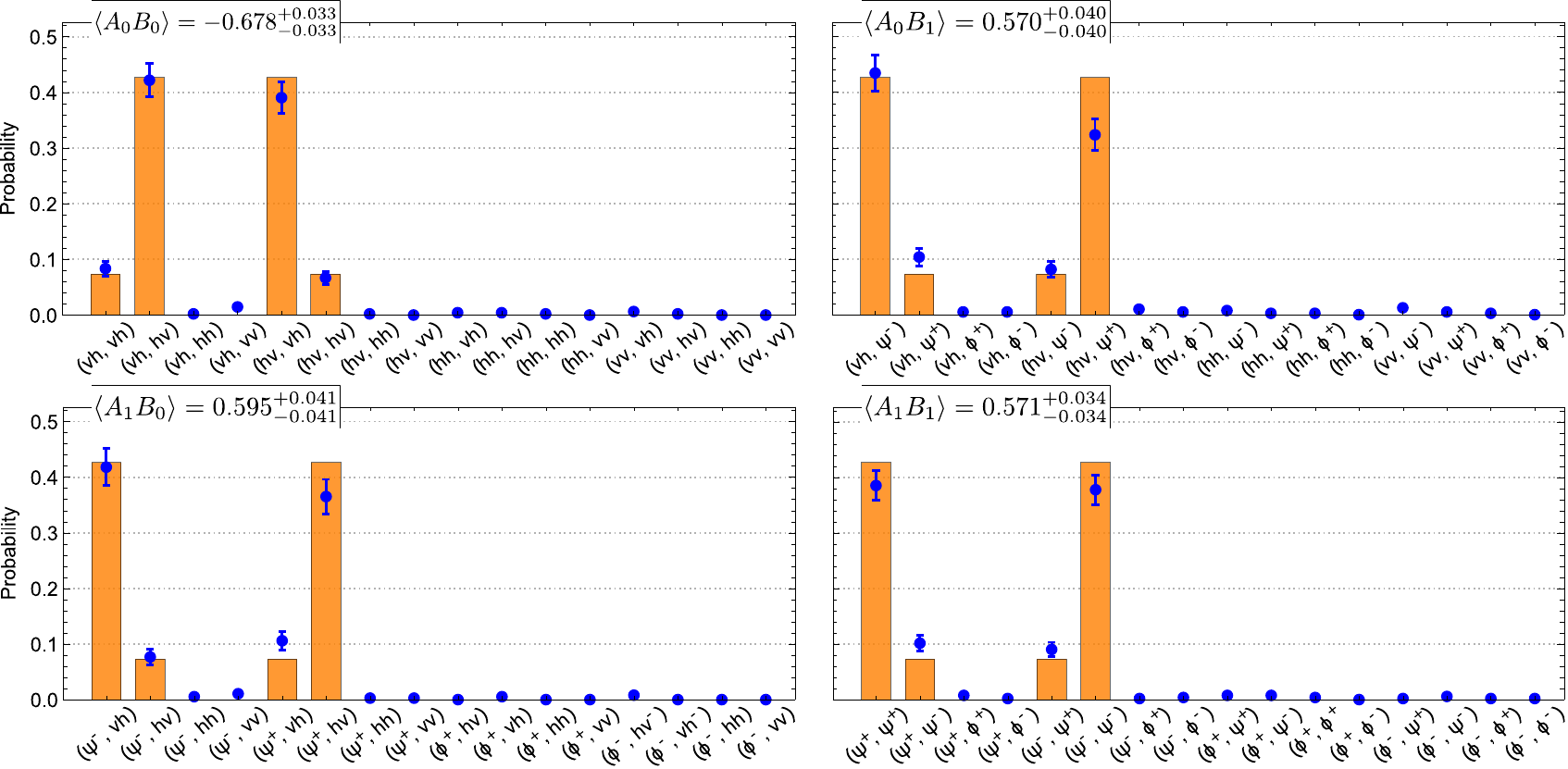}
  \end{center}
\caption{\textbf{Full experimental data}. The full experimental set of probabilities for the 64 settings is shown. The horizontal axis in each of the four plots indicates the eigenstates $(\varphi_A, \varphi_B)$ on which the experimental state shared by Alice and Bob in Eq.~\eqref{eq:4photon_state} is projected, where $\varphi_A$ corresponds to Alice's projection in the two modes $a$ and $\alpha$, $\varphi_B$ instead represents Bob's projection in modes $b$ and $\beta$. For each setting, the number of 6-photon coincidences is recorded and normalised to obtain the relative probabilities as shown in the vertical axis.}
\label{fig:Results64settings}
\end{figure*}

\bigskip
\paragraph*{Measurement protocol.--- }
We now describe in detail the measurement procedure sketched in Fig.~\ref{fig:Setup}. Source $S_0$ and the HWP on its right output arm produce an entangled pair of photons in the state of Eq.~\eqref{Eq:ResourceState}. This photon pair is distributed to the laboratories of Alice's friend and Bob's friend, who measure their photon using Type-I fusion gates~\cite{Browne2005}. Each fusion gate is implemented with a PBS, where horizontally and vertically polarised photons are transmitted and reflected, respectively (by convention collecting a phase $i$ for the latter). Two photons entering the PBS from two different inputs with opposite polarisation, $\ket{h}\ket{v}$ or $\ket{v}\ket{h}$, will exit from the same output port, and will therefore not lead to coincident detection. Only the coincident $\ket{h}\ket{h}$ and $\ket{v}\ket{v}$ components will be recorded in post-selection. For these post-selected photons, the fusion gate induces the following transformations:   
\begin{align}
\ket{h}\ket{h} & \xrightarrow{\text{PBS}} \ket{h}\ket{h} \xrightarrow{\text{Q/HWP}} \ket{h}\frac{\ket{h}+i\ket{v}}{\sqrt{2}}, \notag \\
\ket{v}\ket{v} & \xrightarrow{\text{PBS}} -\ket{v}\ket{v} \xrightarrow{\text{Q/HWP}} -\ket{v}\frac{\ket{h}-i\ket{v}}{\sqrt{2}},
\end{align}
where Q/HWP refers to the combination of a quarter-wave plate at $\pi/4$ and a half-wave plate at $\pi/8$ behind the PBS (see Fig.~\ref{fig:Setup}). The second (heralding) photon in the above equation is then projected onto the state $\ket{h}$ via another PBS. The Type-I fusion gate thus implements the operation
\begin{equation}
FG_\text{I} = \frac{1}{\sqrt{2}} \big( \ket{h}\bra{h}\bra{h} - \ket{v}\bra{v}\bra{v} \big),
\end{equation}
where the factor $\frac{1}{\sqrt{2}}$ indicates the success probability of the gate of $\frac 1 2$.

To use the fusion gate to measure photon $a$ (see Fig.~\ref{fig:Setup}) non-destructively, Alice's friend uses an ancilla from the entangled pair created by $S_A$, prepared as $\ket{\Psi^-}_{\alpha'\alpha}$. Depending on the state of the incoming photon, the operation performed by Alice's friend transforms the overall state as
\begin{align}
\ket{h}_a\ket{\Psi^-}_{\alpha'\alpha} = & \frac{1}{\sqrt{2}} \big( \ket{h}_a\ket{h}_{\alpha'}\ket{v}_{\alpha} - \ket{h}_a\ket{v}_{\alpha'}\ket{h}_{\alpha} \big) \nonumber\\
&\xrightarrow{FG_\text{I}} \frac12 \ket{h}_a\ket{v}_{\alpha}, \notag \\
\ket{v}_a\ket{\Psi^-}_{\alpha'\alpha} = & \frac{1}{\sqrt{2}} \big( \ket{v}_a\ket{h}_{\alpha'}\ket{v}_{\alpha} - \ket{v}_a\ket{v}_{\alpha'}\ket{h}_{\alpha} \big) \nonumber \\
&\xrightarrow{FG_\text{I}} \frac12 \ket{v}_a\ket{h}_{\alpha}.
\label{eq:transfo_FGI}
\end{align}
Hence, the state $\ket{h}_a$ or $\ket{v}_a$ of the external photon in mode $a$ is copied, after being flipped ($h \leftrightarrow v$), onto Alice's friend's photon in mode $\alpha$. In other words, this corresponds to a measurement of the incoming photon in the $\{h,v\}$-basis, with the outcome being recorded in the state of photon $\alpha$, such that we can write
\begin{align}
&\ket{\text{``photon is \textit{h}''}}_\alpha = \ket{v}_\alpha , \nonumber\\
&\ket{\text{``photon is \textit{v}''}}_\alpha = \ket{h}_\alpha. \label{eq:memory_encoding}
\end{align}
The amplitudes $\frac12$ in Eq.~\eqref{eq:transfo_FGI} indicate the total success probability of $\frac14$ for this procedure.

Consider now the central source $S_0$ together with Alice's and Bob's friends' laboratories.
According to Eq.~\eqref{Eq:ResourceState}, the state generated by $S_0$ is, after the unitary $U_{\frac{7\pi}{16}}$,
\begin{align}
\ket{\tilde\Psi}_{ab} = & \frac{1}{\sqrt{2}} \cos \frac{\pi}{8} \, \big(\ket{h}_a\ket{v}_b+\ket{v}_a\ket{h}_b\big) \nonumber\\
& + \frac{1}{\sqrt{2}} \sin \frac{\pi}{8} \, \big(\ket{h}_a\ket{h}_b-\ket{v}_a\ket{v}_b\big).
\end{align}
The transformations induced by Alice's and Bob's friends are then, according to Eq.~\eqref{eq:transfo_FGI}:
\begin{align}
\ket{\tilde\Psi}_{ab}\ket{\Psi^-}_{\alpha'\alpha}\ket{\Psi^-}_{\beta'\beta} \xrightarrow{FG_\text{I}^{\otimes 2}} \frac14 \ket{\tilde \Psi'}_{a \alpha b \beta}, \label{eq:4photon_state_transfo}
\end{align}
with a global success probability of $\frac{1}{16}$. The state
\begin{align}
\ket{\tilde \Psi'}_{a \alpha b \beta} 
& = \frac{1}{\sqrt{2}} \cos \frac{\pi}{8} \, \big(\ket{hv}_{a \alpha}\ket{vh}_{b \beta}+ \ket{vh}_{a \alpha}\ket{hv}_{b \beta}\big) \notag \\
& \ + \frac{1}{\sqrt{2}} \sin \frac{\pi}{8} \, \big( \ket{hv}_{a \alpha} \ket{hv}_{b \beta}-\ket{vh}_{a \alpha} \ket{vh}_{b \beta}\big), 
\label{eq:4photon_state}
\end{align}
is the four-photon state shared by Alice and Bob when both fusion gates are successful.

Recalling from Eq.~\eqref{eq:memory_encoding} how the friends' measurement results are encoded in their polarisation states, the observables of Eq.~\eqref{Eq:Observables} to be measured on $\ket{\tilde \Psi'}_{a \alpha b \beta}$ are
\begin{align}
&A_0 = B_0 = \id \otimes (\ketbra{v}-\ketbra{h}), \nonumber\\
&A_1 = B_1 = \ketbra{\Psi^+} - \ketbra{\Psi^-},
\label{Eq:Observables_v2}
\end{align}
with $\ket{\Psi^\pm} = \frac{1}{\sqrt{2}}(\ket{hv}\pm\ket{vh})$.
To obtain $\langle A_xB_y \rangle$ we project these states onto all combinations of eigenstates of $A_x$ and $B_y$ individually and record 6-photon coincidence events for a  fixed duration. More specifically, to measure $A_0$ (similarly $B_0$) we project onto $\ket{hv}_{a \alpha}$ and $\ket{vv}_{a \alpha}$ (eigenvalue $+1$), and $\ket{hh}_{a \alpha}$ and $\ket{vh}_{a \alpha}$ (eigenvalue $-1$) using a QWP and HWP to implement local rotations before the final PBS, not using the BS in Fig.~\ref{fig:Setup}. Note that $A_0$ cannot be simply measured by ignoring photon $a$, due to the probabilistic nature of the photon source. Hence, this photon has to be measured in a polarisation-insensitive way, which, due to the polarisation-sensitive nature of the photon-detectors, is best achieved by summing over the projections onto both orthogonal polarisations.
To measure $A_1$ ($B_1$) we use a $50/50$ beam splitter followed by projection onto $\ket{vh}$. Due to nonclassical interference in the beam splitter, this implements a projection onto the singlet state $\ket{\Psi^-}_{a \alpha}$ with success probability $\frac 1 2$. Using quantum measurement tomography, we verified this Bell-state measurement with a fidelity of $\mathcal{F}_{\textsc{bsm}}=96.84^{+0.05}_{-0.05}$. Projections on the other Bell states are possible via local rotations using the same QWP and HWP as before. Here $\ket{\Psi^+}_{a \alpha}$ takes eigenvalue $+1$, $\ket{\Psi^-}_{a \alpha}$ eigenvalue $-1$, and $\ket{\Phi^\pm}_{a \alpha}= \frac{1}{\sqrt{2}}(\ket{hh}\pm\ket{vv})_{a \alpha}$ eigenvalue $0$.
Probabilities are obtained from normalising the measured counts with respect to the total of the 16 measurements for each pair of observables, see Fig.~\ref{fig:Results64settings}. The theoretically expected values for the various probabilities are either $\frac{1}{4}(1+\frac{1}{\sqrt{2}}) \simeq 0.427$, $\frac{1}{4}(1-\frac{1}{\sqrt{2}}) \simeq 0.073$, or $0$. In addition to this result, an alternative measurement protocol for $A_0$ and $B_0$ is presented below. 

\bigskip

\paragraph*{Error analysis.--- }
As described previously, each average value $\langle A_{x}B_{y} \rangle$ is calculated from 16 measured 6-fold coincidence counts $n_i$. These numbers follow a Poisson distribution with variance $\sigma_{n_i}^2 = n_i$. The uncertainty on $\langle A_{x}B_{y} \rangle = f(n_1,\ldots,n_{16})$ can then be computed using
\begin{equation}
\sigma^{2}_{f}(n_{1},\ldots,n_{16})=\sum_{i=1}^{16}\left(\frac{\partial f}{\partial n_{i}}\right)^{2}\sigma_{n_{i}}^{2}  .
\end{equation}
Since the four averages $\langle A_{1}B_{1} \rangle$, $\langle A_{1}B_{0} \rangle$, $\langle A_{0}B_{1} \rangle$ and $\langle A_{0}B_{0} \rangle$ are statistically independent, the uncertainties can be calculated independently and combined to estimate the uncertainty on $S$. To take into account potentially asymmetric errors in the limit of small count rates, we computed the uncertainty on the Bell-Wigner parameter $S$ using a Monte-Carlo routine with {100 000} samples. The values obtained through these two methods agree to within $0.0032$.

\begin{figure}[t!]
  \begin{center}
\includegraphics[width=0.95\columnwidth]{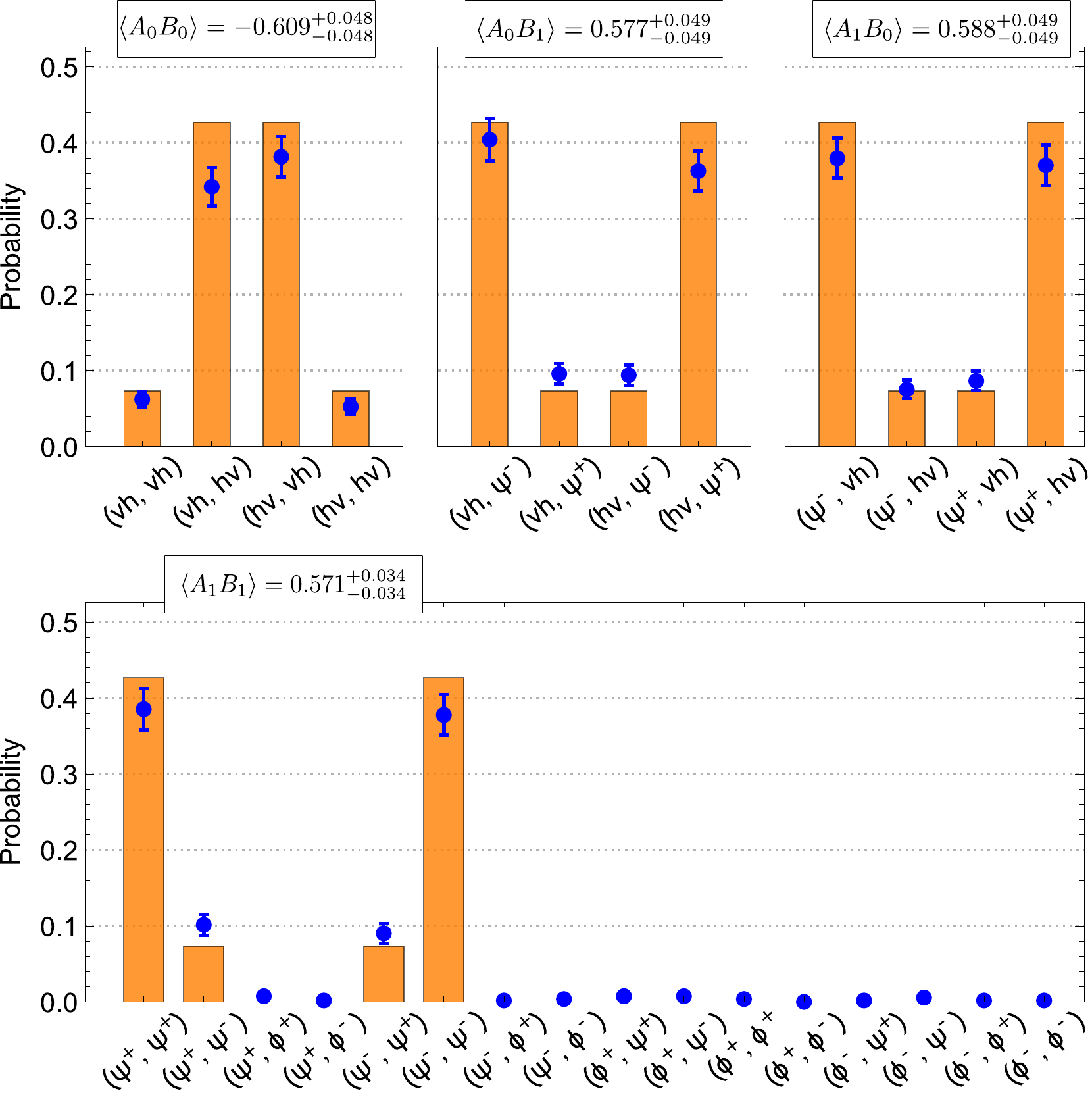}
  \end{center}
\caption{\textbf{Alternative protocol experimental data}. The experimental probabilities obtained with the alternative definition of $A_0$ and $B_0$, Eq.~\eqref{Eq:Observables_Caslav}, are shown. $\expec{A_1B_1}$, in the bottom panel, is left unchanged by the new definition thus the data shown here as well as the average value for this couple of observables, is the same as in Figs.~\ref{fig:Results} and~\ref{fig:Results64settings}. $\expec{A_0B_0}$, $\expec{A_0B_1}$ and $\expec{A_1B_0}$ shown in the top panels are instead measured in accordance with the new protocol. In this case, only 6-photon coincidences for the non-zero terms, labelled in the horizontal axis, are recorded and normalised with the sum of all the coincidences recorded for $\expec{A_1B_1}$.}
  \label{fig:ResultsPOL}
\end{figure}

Note that in the results shown in Fig.~\ref{fig:ResultsPOL} with the observables of Eq.~\eqref{Eq:Observables_Caslav}, errors are correlated due to normalisation with a common total. Accounting for this in the error propagation results in slightly larger statistical uncertainty.

The Bell-Wigner value $S_{exp}$ that can be achieved experimentally is primarily limited by multi-pair emissions from our probabilistic photon-pair sources. We first note that any emission of 3 pairs from any subset of our 3 sources occurs with roughly similar probability. To exclude unwanted terms we use six-fold coincidence detection, which can only be successful for an emission of one pair each in $S_0$, $S_A$ and $S_B$, or three pairs in $S_0$. The latter would amount to noise but is excluded by our cross-polarisation design and can thus not lead to a coincidence detection. This leaves higher-order contributions where at least 4 photon pairs are produced as the main source of errors. Since such events scale with a higher exponent of the pump power, they are suppressed in our experiment by working with a relatively low pump power of 100 mW.

\bigskip
\paragraph*{Towards a loophole-free ``Bell-Wigner'' test.--- } 
Since our experiment relies on some of the same assumptions as traditional Bell tests, it is subject to the same conceptual and technical loopholes: locality, freedom of choice, and the detection loophole. Due to the increased complexity of our experiment, compared to a standard Bell test, the practical requirements for closing these loopholes are significantly more challenging. We now briefly discuss how these loopholes could be closed in the future.

The configuration of our experiment makes it analogous to an ``event-ready'' Bell test, where the detection of the ancilla photons in the fusion gates heralds which events should be kept for the Bell-Wigner test. In such a configuration, closing the locality and freedom of choice loopholes requires the heralding events to be space-like separated from Alice's and Bob's setting choices, which should each be space-like separated from the measurement outcome of the other party. This imposes stringent space-time location requirements for a Bell-Wigner test closing these loopholes.

The detection loophole arises because only a fraction of all created photons is detected. In our ``event-ready'' configuration, the limited success probability of the fusion gates is not an issue: only heralded events will contribute to the Bell-Wigner test. Nevertheless, to ensure that the fusion gates are indeed event-ready, the ancilla detectors should be photon-number-resolving. 

To measure the observables $A_x, B_y$, we chose to project the photon states onto their different eigenstates separately. To close the detection loophole one cannot follow such an approach: the measurement protocol should be able to project the states onto all of the eigenstates in any run of the experiment.

To measure $A_0/B_0$ from Eq.~\eqref{Eq:Observables}, one could pass the friend's photon through a PBS, with detectors at both outputs. As for $A_1/B_1$, a full Bell-state measurement (which is impossible with linear quantum optics~\cite{Calsamiglia2001}) is not required: it suffices to distinguish $\ket{\Psi^+}, \ket{\Psi^-}$, and have a third outcome for $\ket{\Phi^\pm}$ (see Eq.~\eqref{Eq:Observables_v2}). This can be realised with a small modification to our setup, with detectors added on the second outputs of Alice's and Bob's PBS~\cite{Braunstein1995}. An even simpler measurement would discriminate e.g.\ $\ket{\Psi^-}$ from the other three Bell states, thus measuring the observables $A_1 = B_1 = \id - 2 \ketbra{\Psi^-}$; this would not change anything in an ideal implementation, but simplifies the analysis with detection inefficiencies below.

Even the best photon detectors aren't 100\% efficient and optical loss is unavoidable. Assuming a symmetric combined detection efficiency per photon of $\eta$, the measurement of $A_0/B_0$ requires one detector to click and would succeed with probability $\eta$, while the measurement of $A_1/B_1$ requires two detectors to fire and would work as expected with probability $\eta^2$. When a detector fails to click, a simple strategy is to output a fixed pre-defined value for the measurement outcome, e.g.\ $+1$. Then, for Eq.~\eqref{eq:4photon_state} the average values $\langle A_{x}B_{y} \rangle$ are theoretically expected to be $\langle A_{0}B_{0} \rangle = \eta^2(-\frac{1}{\sqrt{2}}) + (1-\eta)^2$, $\langle A_{0}B_{1} \rangle = \langle A_{1}B_{0} \rangle = \eta^3\frac{1}{\sqrt{2}} + (1-\eta)(1-\eta^2)$ and $\langle A_{1}B_{1} \rangle = \eta^4\frac{1}{\sqrt{2}} + (1-\eta^2)^2$. With these values, the minimal required detection efficiency to violate inequality~\eqref{eq:CHSH} with (unrealistically) perfect quantum states and measurements is $\eta > 2\sqrt{3(1-\frac{1}{\sqrt{2}})}-1 \simeq 0.875$. This is a more stringent requirement than for a standard test of the CHSH inequality, for which a similar analysis for maximally entangled states yields $\eta > 2\sqrt{2}-2 \simeq 0.828$. To relax this requirement, one might attempt similar tricks as for standard Bell tests, e.g.\ to use non-maximally entangled states~\cite{Eberhard1993}, although this will come at the cost of a reduced violation of the inequality.

Note, finally, that in the conclusions we draw from the violation of inequality~\eqref{eq:CHSH}, we need to trust that $A_0$ and $B_0$ indeed directly measure the memory of Alice's and Bob's friends, so as to unveil their respective facts. A new loophole may be opened, now specific to Bell-Wigner tests, if such an interpretation cannot be maintained. To address this loophole with a setup like ours, one should use measurement devices for $A_0$ and $B_0$ that clearly separate the initial systems and the memories of each friend, and only ``looks'' at the memory photons, rather than at the system photon + memory photon together; we also leave this possibility as a challenge for future Bell-Wigner experimental tests.

\bigskip
\paragraph*{Alternative observables $A_0, B_0$.--- }

In Ref.~\cite{Brukner2018} the observables $A_0, B_0$ were defined as
\begin{align}
A_0 = B_0 & = \ \ketbra{h}\!\otimes\!\ketbra{\text{``photon is \textit{h}''}} \notag \\
& \quad - \ketbra{v}\!\otimes\!\ketbra{\text{``photon is \textit{v}''}}, \label{Eq:Observables_Caslav}
\end{align}
which have a slightly different physical interpretation. The observables used in the main text and defined in Eq.~\eqref{Eq:Observables}, directly measure the facts established by the friend, as recorded in their memory. In contrast, the observables in Eq.~\eqref{Eq:Observables_Caslav} can be understood as not only a measurement of the friend's record (to establish a ``fact for the friend''), but also of the original photon measured by the friend, as a consistency check: if the state of the photon is found to be inconsistent with the friend's record, the definition above assigns a value $0$ for the measurement result.

Our experiment also allows us to test inequality~\eqref{eq:CHSH} using this alternative definition of $A_0, B_0$. Indeed, from the experimental data shown in Fig~\ref{fig:Results64settings}, it suffices (according to Eq.~\eqref{Eq:Observables_Caslav} and recalling Eq.~\eqref{eq:memory_encoding}) to assign the eigenstate/eigenvalue according to $\ket{hv} \to +1$, $\ket{vh} \to -1$ and $\ket{hh}, \ket{vv} \to 0$ in the calculation of the average values $\langle A_x B_y \rangle$.
We thus obtain the three average values $\langle A_0 B_0 \rangle = 0.662^{+0.033}_{-0.033}$, $\langle A_0 B_1 \rangle = 0.573^{+0.039}_{-0.039}$ and $\langle A_1 B_0 \rangle = 0.600^{+0.040}_{-0.040}$ with $\langle A_1 B_1 \rangle$ unchanged. With these values, we have $S_{exp} = 2.407^{+0.073}_{-0.073}$, again violating inequality~\eqref{eq:CHSH} by more than 5 standard deviations. As in the main text, errors are computed assuming Poissonian photon counting statistics, see below for details.

\bigskip
\paragraph*{Alternative measurement protocol for $A_0, B_0$.--- }
Recall that in order to measure $A_0$ (similarly $B_0$), the beam splitter for Alice in Fig.~\ref{fig:Setup} has to be removed relative to the measurement of $A_1$. A less invasive method (which does not compromise the alignment of our optical elements) is to introduce linear polarisers in modes $a (b)$ and $\alpha (\beta)$. This effectively measures the photons before the BS, preventing interference.

We implemented this procedure for the alternative definition of $A_0$ and $B_0$ in Eq.~\eqref{Eq:Observables_Caslav}. Since this approach leads to a reduced success probability of the measurement of $A_0 (B_0)$ by a factor $1/4$, we measured all 16 eigenvectors only for $\expec{A_1B_1}$. For the other observables we measured the eigenvectors with non-zero eigenvalues and normalised all data with respect to the total counts for $\expec{A_1B_1}$, Fig.~\ref{fig:ResultsPOL}. This slightly increases experimental uncertainties, which we have taken into account in our error analysis. The expectation values so obtained are $\langle A_0 B_0 \rangle = 0.609^{+0.048}_{-0.048}$, $\langle A_0 B_1 \rangle = 0.577^{+0.049}_{-0.049}$ and $\langle A_1 B_0 \rangle = 0.588^{+0.049}_{-0.049}$ with $\langle A_1 B_1 \rangle$ unchanged, and $S_{exp}=2.346^{+0.110}_{-0.110}$, violating the Bell-Wigner inequality by more than 3 standard deviations. We note that the violation observed with this method is somewhat reduced because of $\sim 4.83\pm 0.97\%$ loss that is introduced by the polarisers. This effectively reduces the number of counts that are observed in the settings $A_0$ and $B_0$ compared to the normalisation used, and thereby reduces the expectation values $\langle A_0 B_1 \rangle$ and $\langle A_1 B_0 \rangle$, and $\langle A_0 B_0 \rangle$, leading to a reduced violation.

\end{document}